\documentclass[useAMS,usedcolumn]{mn2e}
\usepackage{supertabular,epsf,rotate,graphicx}
\newcommand{\f}{\phantom{2}}
\newcommand{\mc}{\multicolumn}

\newcommand{\ltsimeq}{\raisebox{-0.6ex}{$\,\stackrel 
        {\raisebox{-.2ex}{$\textstyle <$}}{\sim}\,$}} 

\def\chandra{{\it Chandra~}}
\def\xmm{{\it XMM-Newton~}}

\begin{document}

\title[Obscured AGN from the EDXS] {Obscured AGN from the ELAIS Deep
X-ray Survey}

\author[Willott et al.]
{C.J.\ Willott$^{1,2}$\footnotemark,
C.\ Simpson$^{3}$,
O.\ Almaini$^{4}$, 
J.C.\ Manners$^{4}$, 
O.\ Johnson$^{4}$,\and
A.\ Lawrence$^{4}$,
J.S.\ Dunlop$^{4}$, 
R.J.\ Ivison$^{5}$,
S.\ Rawlings$^{2}$,
E.\ Gonz\'alez-Solares$^{6,8}$,\and
I.\ P\'erez-Fournon$^{6}$,
S.\ Serjeant$^{7}$,
S.J.\ Oliver$^{8}$,
N.D.\ Roche$^{4}$,
R.G.\ Mann$^{4}$,\and
M.\ Rowan-Robinson$^{9}$\\
$^1$Herzberg Institute of Astrophysics, National Research Council, 
5071 West Saanich Rd, Victoria, B.C. V9E 2E7, Canada\\
$^2$Astrophysics, Department of Physics, Keble Road, Oxford, OX1
3RH, U.K. \\
$^3$Subaru Telescope, National Astronomical Observatory of Japan, 650
N. A`oh\={o}k\={u} Place, University Park, Hilo, Hawaii, 96720, USA \\
$^4$Institute for Astronomy, University of Edinburgh, Royal
Observatory,  Blackford Hill, Edinburgh EH9 3HJ\\
$^5$UK Astronomy Technology Centre, Royal Observatory, Blackford Hill, Edinburgh, EH9 3HJ\\
$^6$Instituto de Astrof\'\i sica de Canarias, C/ Via
Lactea s/n, 38200 La Laguna, Tenerife, Spain \\
$^7$Unit for Space Sciences and Astrophysics, School of Physical
Sciences, University of Kent, Canterbury, CT2 7NZ\\
$^8$Astronomy Centre, CPES, University of Sussex, Falmer, Brighton, 
BN1 9QJ\\
$^9$Astrophysics Group, Blackett Laboratory, Imperial College,
Prince Consort Rd., London, SW7 2BW\\}

\maketitle

\begin{abstract}

The sources discovered in deep hard X-ray surveys with 2-8 keV fluxes
of ~~~~~~ $S_{2-8}\sim10^{-14}$\,erg\,cm$^{-2}$\,s$^{-1}$ make up the
bulk of the X-ray background at these energies. We present here
detailed multi-wavelength observations of three such sources from the
ELAIS Deep X-ray Survey. The observations include sensitive
near-infrared spectroscopy with the Subaru Telescope and X-ray
spectral information from the {\it Chandra X-ray Observatory}. The
sources observed all have optical-to-near-IR colours redder than an
unobscured quasar and comprise a reddened quasar, a radio galaxy and
an optically-obscured AGN. The reddened quasar is at a
redshift $z=2.61$ and shows a very large X-ray absorbing column of
$N_{\rm H} \approx 3 \times 10^{23} {\rm cm}^{-2}$. This contrasts
with the relatively small amount of dust reddening, implying a
gas-to-dust ratio along the line-of-sight a hundred times greater than
that of the Milky Way. The radio galaxy at $z=1.57$ shows only narrow
emission lines, but has a surprisingly soft X-ray spectrum. The
softness of this spectrum either indicates an unusually low
gas-to-dust ratio for the absorbing medium or X-ray emission related
to the young radio source. The host galaxy is extremely red
($R-K=6.4$) and its optical/near-IR spectrum is best fit by a strongly
reddened ($A_V \approx 2$) starburst. The third X-ray source discussed
is also extremely red ($R-K=6.1$) and lies in a close grouping of
three other $R-K>6$ galaxies. No emission or absorption lines were
detected from this object, but its redshift (and that of one of the
nearby galaxies) are constrained by SED-fitting to be just greater
than $z=1$. The extremely red colours of these two galaxies can be
accounted for by old stellar populations. These observations
illustrate the diverse properties of hard X-ray selected AGN at high
redshift in terms of obscuration at optical and X-ray wavelengths and
the evolutionary states of their host galaxies.

\end{abstract}


\footnotetext{Email: chris.willott@nrc.ca}

\section{Introduction}

Deep extragalactic surveys with the \chandra and XMM observatories are
capable of resolving most of the hard (2-8 keV) X-ray background
(e.g. Cowie et al. 2002). Early results from such surveys
(e.g. Hornschemeier et al. 2001; Tozzi et al. 2001), confirm that
almost all of the X-ray background is produced by accretion of
material onto supermassive black holes in active galactic nuclei
(AGN). The hard spectrum of the X-ray background and the fact that the
majority of hard X-ray sources do not have quasar counterparts shows
that obscuration is extremely important. One of the key goals of the
current generation of X-ray surveys is understanding the nature and
evolution of the gas and dust surrounding AGN and the reprocessing
effects they have upon the emitted radiation.

The ELAIS Deep X-ray Survey (EDXS) is a deep survey with the \chandra
Observatory in two fields which have been well-studied at other
wavebands. In each of the fields, designated N1 and N2, \chandra
ACIS-I observations of duration 75 ks have been made. A total of 225
X-ray sources are detected in the full band images above a flux limit
of $S_{0.5-8{\rm keV}}>1.1 \times 10^{-15}~ {\rm erg ~cm}^{-2}~{\rm
s}^{-1}$ (Manners et al. 2002). Details of the optical identifications
of the X-ray sources are presented in Gonz\'alez-Solares et
al. (2002). These fields lie within the European Large-Area ISO Survey
(ELAIS) and have been observed with ISO at 7, 15, 90 and 175 $\mu$m
(Oliver et al. 2000) and with the Very Large Array (VLA) at 1.4-GHz
(Ciliegi et al. 1999). One of the two X-ray fields is coincident with
the widest survey yet made with SCUBA on the JCMT (Scott et al. 2002)
and a very deep VLA map (Ivison et al. 2002). The relationship between
the X-ray and sub-mm sources is discussed in Almaini et al. (2002).

With a median $2-8$ keV flux of $S_{2-8}=6 \times 10^{-15} $\,erg\,cm$
^{-2}$\,s$^{-1}$, the hard X-ray sources present in the EDXS \chandra
data are close to the peak of the source contribution to the X-ray
background at these energies (Cowie et al. 2002).  Determining the
nature of these sources is therefore an essential part of
understanding the X-ray background. In this paper, we present detailed
observations of three hard X-ray sources selected from the \chandra
observations of the EDXS. These sources are all considerably redder
than an unobscured quasar and comprise a reddened quasar, a radio
galaxy and a totally obscured, Type 2, radio-quiet AGN. The latter two
sources are optically faint ($R>25$) and extremely red
($R-K>6$). Table \ref{tab:sample} shows the basic information on the
\chandra sources discussed in this paper.

In Section 2 we present details of the spectroscopic observations
performed with the Subaru Telescope and the William Herschel
Telescope. In Sections 3-5 we discuss each object in detail. The
conclusions are given in Section 6. We assume throughout that $H_0=70~
{\rm km~s^{-1}Mpc^{-1}}$, $\Omega_{\mathrm M}=0.3$ and
$\Omega_\Lambda=0.7$. The convention for all spectral indices,
$\alpha$, is that $S_{\nu} \propto \nu^{-\alpha}$,
where $S_{\nu}$ is the flux-density at frequency $\nu$.

\begin{table*}
\footnotesize
\begin{center}
\begin{tabular}{ccccrrcc}
\hline
\mc{1}{l}{Source} &\mc{1}{c}{X-ray position} &\mc{1}{c}{Near-IR position} &\mc{1}{c}{X-ray flux (0.5-8 keV)}&\mc{1}{c}{HR} &\mc{1}{c}{$R$ mag} &\mc{1}{c}{$K$ mag} \\
\mc{1}{c}{ } &\mc{1}{c}{(J2000.0)} &\mc{1}{c}{(J2000.0)} &\mc{1}{c}{(erg cm$^{-2}$ s$^{-1}$)} &\mc{1}{c}{} &\mc{1}{c}{} &\mc{1}{c}{} \\

\hline

N2\_21 & 16:36:58.07 +40:58:21.1 & 16:36:58.05 +40:58:20.6 & $(21.5 \pm 1.6)\times 10^{-15}$ & $-0.42 \pm 0.07$ & $25.02 \pm 0.09$ & $18.60 \pm 0.08$\\
N2\_25 & 16:36:55.79 +40:59:10.5 & 16:36:55.79 +40:59:10.4 & $(10.2 \pm 1.1)\times 10^{-15}$ & $ 0.26 \pm 0.11$ & $23.07 \pm 0.03$ & $19.10 \pm 0.10$\\
N2\_28 & 16:36:55.21 +40:59:44.1 & 16:36:55.25 +40:59:44.2 &$\f (3.8 \pm 0.7)\times 10^{-15}$& $-0.14 \pm 0.18$ & $25.81 \pm 0.21$ & $19.71 \pm 0.05$\\

\hline
\end{tabular}

{\caption[junk]{\label{tab:sample} ELAIS Deep X-ray Survey \chandra
sources observed with Subaru near-infrared spectroscopy. The first
column gives the informal name of the source (official IAU
designations are given in Section \ref{nirspec}). The second and third
columns are the X-ray and near-IR positions respectively. The
offsets between these positions are $<0.5$ arcsec in all cases. The
fourth column gives the X-ray flux in the full (0.5-8 keV) band and
the fifth column the hardness ratio HR=(H-S)/(H+S) (so that harder
sources have higher values of HR; see Manners et al. 2002). The final
two columns are optical and near-IR magnitudes measured in a 3 arcsec
aperture. The imaging data are described in Gonz\'alez-Solares et
al. (2002), Roche et al. (2002) and Ivison et al. (2002).}}  \normalsize \end{center}
\end{table*}

\section{Observations}

\subsection{Subaru Near-IR Spectroscopy}
\label{nirspec}

We selected suitable targets for near-IR spectroscopy from the EDXS N2
region according to the following criteria: (i) detected in the hard
(2-8 keV) band; (ii) optically-faint, $R>25$; (iii) extremely red colours,
$R-K>6$. Two sources meeting these criteria, N2\_21
(CXOEN2~J163658.0+405821) and N2\_28 (CXOEN2~J163655.2+405944), were
observed with the OH-airglow suppression Spectrograph (OHS; Iwamuro et
al. 2001) on the Subaru Telescope on the night of UT 2001 June
11. This spectrograph uses a fixed grism giving simultaneous $J$ and
$H$-band spectra over the wavelength ranges $1.108-1.353 \umu$m and
$1.477-1.804 \umu$m. The seeing was $0.7-0.8$ arcsec and a 0.95 arcsec
slit was used giving a resolution of $\approx 80$ \AA~ (equivalent to
$\approx 1500$ km s$^{-1}$). The total exposure times per object were
4000 s, split into 4 nodded frames of 1000 s each. The F8 star SAO
46272 was observed immediately after each target to enable correction
for atmospheric extinction.

The OHS data were reduced in a standard manner of flat-fielding, sky
subtraction, registration and co-addition with cosmic ray
rejection. Since the OHS $JH$ grism is fixed, there was no external
wavelength calibration and the polynomial fit given on the OHS web
page was used for wavelength calibration. Relative flux-calibration
and atmospheric extinction corrections were applied using the F8
star. Photometric flux-calibration was performed by scaling the
reduced spectra by the fluxes measured in a 3 arcsec aperture from our
own near-IR imaging (Gonz\'alez-Solares et al. 2002).  We checked that
any emission lines have similar spatial extension to the continuum and
therefore this method does not over- or under-estimate the emission
line fluxes.

Optical spectroscopy of the X-ray source N2\_25
(CXOEN2~J163655.7+405910) had already revealed it to have a redshift
of $z=2.61$ (see Section \ref{optspec}) and properties resembling a
reddened quasar. Therefore we decided to obtain a spectrum of N2\_25
in the $K$-band, since this contains the H$\alpha$ emission line.
N2\_25 was observed with the Infrared Camera and Spectrograph (IRCS;
Kobayashi et al.\ 2000) on Subaru Telescope on the night of UT 2001
May 5. The seeing was $0.5-0.6$ arcsec. The $K$-grism was used with
the 58\,mas pixel scale and a 0.6-arcsec slit (aligned east--west) to
give a spectral resolution of $\sim 600$\,km\,s$^{-1}$ at the
wavelength of redshifted H$\alpha$.  Eight separate 600-second
exposures were taken, alternating between two positions separated by 7
arcsec along the slit. First-order background subtraction was
accomplished by subtracting pairs of exposures, and the {\footnotesize
IRAF} task {\footnotesize BACKGROUND} was used to remove residual sky
emission. These four background-subtracted images were then combined
with cosmic ray rejection to produce the final 2D image.  Spectra were
extracted with 1-arcsec apertures along both the positive and negative
beams and combined. The spectrum was flux-calibrated and corrected for
atmospheric absorption by ratioing with the F8 star SAO~46107 (which
had been reduced in the same manner), whose flux scale had been tied
to the UKIRT standard FS~132.

\begin{figure*} 
\vspace{-0.5cm} 
\epsfxsize=0.97\textwidth 
\epsfbox{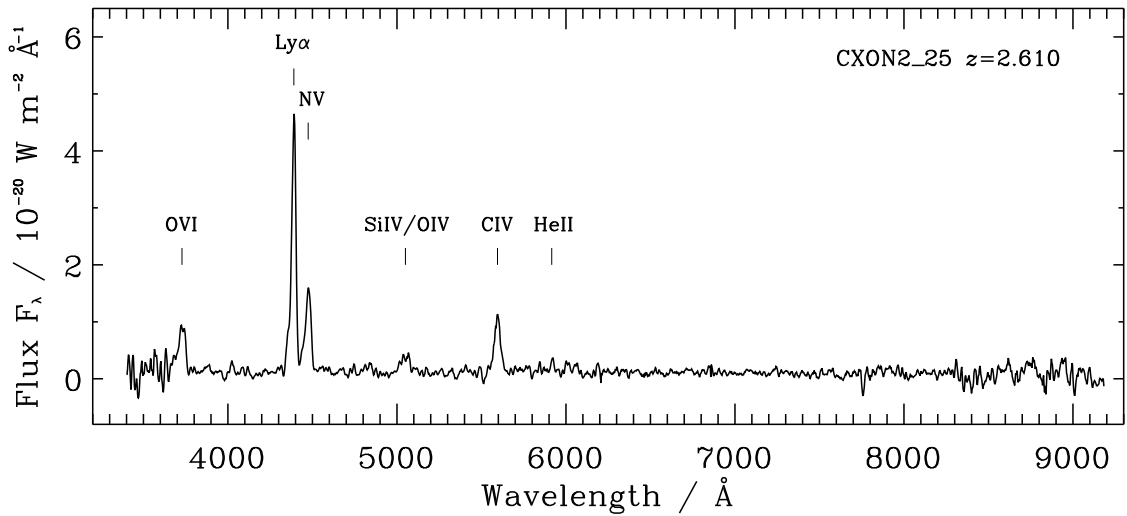} 
\vspace{-0.5cm}
{\caption[junk]{\label{fig:specwht} Optical spectrum of N2\_25 with
emission lines labelled. The spectrum has been smoothed with a boxcar
filter of width 15 \AA.}}
\end{figure*}

\subsection{Optical spectroscopy}
\label{optspec}

These three X-ray sources were also observed using the ISIS
spectrograph at the William Herschel Telescope on 2001 May 18 and
21. The low resolution R158R and R158B gratings were used giving
continuous wavelength coverage from $3000-9200$ \AA. The seeing was
1.0 arcsec. A slit width of 2 arcsec gave a resolution of 12 \AA~
(equivalent to $700$ km s$^{-1}$ at 5000 \AA). N2\_21 and N2\_25 were
observed simultaneously using a slit position angle of $153.2^{\circ}$
for a total integration time of 1800 s. N2\_28 was observed
simultaneously with two other X-ray sources (N2\_32 and N2\_34 --
Willott et al. 2001a) along a position angle of $39.5^{\circ}$ for a
total integration time of 3600 s. Data reduction was performed using
standard procedures which are fully described in Willott et
al. (1998). N2\_25 was clearly detected and is fully discussed in
Section \ref{n225}. Neither continuum nor emission lines were detected
in the optical spectra of N2\_21 or N2\_28.

\section{N2\_25 - a lightly reddened quasar with a very hard X-ray spectrum}
\label{n225} 

\begin{figure*}
\vspace{-0.8cm}
\hspace{-0.7cm}
\epsfxsize=0.97\textwidth \epsfbox{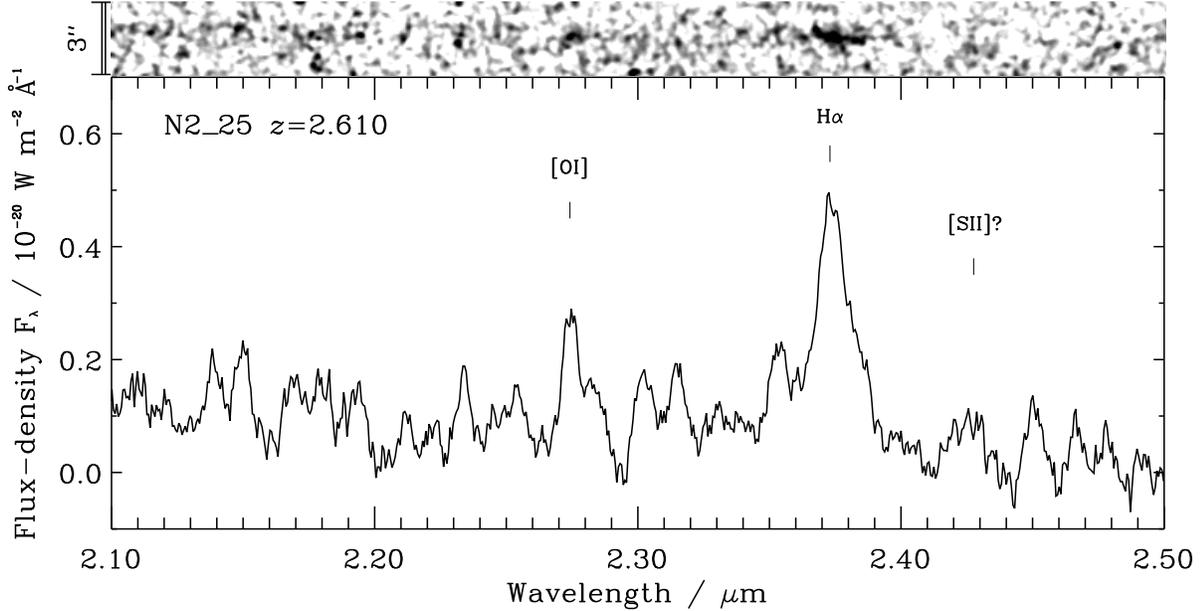}
\vspace{-15.2cm}
{\caption[junk]{\label{fig:specn225} IRCS $K$-band spectrum of N2\_25
with emission lines labelled. A greyscale representation of the
two-dimensional spectrum (3 arcseconds along the slit, centred on the
target) is shown at top. The H$\alpha$ and [O\,{\small I}] lines are
clearly visible in the two-dimensional spectrum. The possible [SII]
emission feature is less reliable since its peak appears to be offset
by 0.4 arcsec from the other two lines.}}
\end{figure*}

\begin{table*}
\footnotesize
\begin{center}
\begin{tabular}{llccr}
\hline
\mc{1}{l}{Emission line} &\mc{1}{c}{$\lambda_{\rm obs}$} &\mc{1}{c}{$z_{\rm em}$} &\mc{1}{c}{FWHM}
&\mc{1}{c}{Flux/$10^{-19}$ }  \\
\mc{1}{c}{ } &\mc{1}{c}{(\AA)} &\mc{1}{c}{ } &\mc{1}{c}{(km s$^{-1}$)} &\mc{1}{c}{(W m$^{-2}$)} \\

\hline
O\,VI             $\lambda \lambda 1032, 1038$ & \f 3731 $\pm$ 5 & 2.606 & ---  &     \f  4.3 $\pm$ 1.0 \\
Ly$\alpha\,                         \lambda 1216$ & \f 4390 $\pm$ 2 & 2.610 & 1980 &          17 $\pm$ 2.0 \\
N\,V                               $\lambda 1240$ & \f 4475 $\pm$ 1 & 2.609 & 2410 &     \f  7.0 $\pm$ 1.3 \\
Si\,IV $\lambda 1397$ + O\,IV      $\lambda 1402$ & \f 5052 $\pm$ 12& 2.609 & ---  &     \f  1.7 $\pm$ 0.6 \\
C\,IV                              $\lambda 1549$ & \f 5595 $\pm$ 2 & 2.612 & 2310 &     \f  4.5 $\pm$ 0.8 \\
He\,II                             $\lambda 1640$ & \f 5920 $\pm$ 2 & 2.610 & ---  &     \f  0.7 $\pm$ 0.4 \\
${\rm [O\,I]}\,                     \lambda 6300$ &   22741 $\pm$ 5 & 2.610 &\f 650&     \f  1.3 $\pm$ 0.4 \\
H$\alpha\, \lambda 6563$ + [N\,II] $\lambda \lambda 6548, 6583$ &23737 $\pm$ 6&2.617&1900&\f 5.8 $\pm$ 0.6 \\
${\rm [S\,II]}\,   \lambda 6716 + \lambda 6731$ ? &   24275 $\pm$ 10& 2.610 & ---  &     \f  0.8 $\pm$ 0.5 \\

\hline
\end{tabular}

{\caption[junk]{\label{tab:linesn225} Emission line data for N2\_25
from the optical and $K$-band spectra. For very weak detections or
blended lines of comparable strength, no FWHM is given. The line
widths have not been deconvolved, but are several times greater than
the instrumental resolution in all cases except for that of the
unresolved [O\,I] line. 

}} 
\normalsize
\end{center}
\end{table*}

The X-ray source N2\_25 is identified optically with an unresolved (in
0.8 arcsec seeing) object with $R=23.07$. The corresponding $K$-band
counterpart is also spatially unresolved. N2\_25 has an
optical/near-infrared colour of $R-K=4.0$, considerably redder than
the colours of typical high-redshift quasars which have
$R-K\approx2.0$.

\subsection{Spectroscopic data}
\label{n225spec}

In Fig.\ \ref{fig:specwht} we show the optical spectrum of N2\_25 in a
$2\times2$ arcsec$^2$ aperture. The spectrum reveals a weak continuum and
several strong, relatively narrow emission lines. The $K$-band
spectrum of N2\_25 is shown in Fig.\ \ref{fig:specn225}. The continuum
is clearly detected and also there are emission lines of H$\alpha$,
[O\,{\small I}] and (marginally) [S\,{\small II}]. Measurements of
the emission lines from these two spectra are given in Table
\ref{tab:linesn225}. The lines are consistent with a redshift of
$z=2.610$.

The H$\alpha$, Ly$\alpha$, N\,{\small V} and C\,{\small IV} lines have
FWHM in the range $1900 - 2400$ km s$^{-1}$. These velocity widths are
close to the traditional dividing line between the broad line region
(BLR) and the narrow line region (NLR). In contrast the only forbidden
line detected, [O\,{\small I}], has a much narrower line width which
is virtually unresolved at a resolution of $600$ km s$^{-1}$. This
suggests that the permitted transitions are from the BLR. Further
support for this inference comes from the very strong N\,{\small V}
emission. N2\_25 has line ratios N\,{\small V} / C\,{\small IV}$ = 1.6
\pm 0.6$ and N\,{\small V} / He\,{\small II}$ = 10 \pm 6$. These
ratios are consistent with those of the BLR of super-solar metallicity
quasars (Hamann \& Ferland 1993) and much greater than the ratios
observed in the NLR of high-redshift radio galaxies (Vernet et
al. 2001).  The narrow emission lines in most high-redshift radio
galaxies and in at least one X-ray selected Type-II quasar are
spatially resolved (Jarvis et al. 2001; Stern et al. 2002).  Therefore
a further test of whether the observed emission lines are from the BLR
or NLR is their spatial extent. Both the Ly$\alpha$ and N\,{\small V}
lines appear marginally resolved with FWHM = 1.2 arcsec, compared to
the continuum FWHM $\approx 1.0$ arcsec. However, the low
signal-to-noise in the continuum hampers an accurate FWHM measurement
and this difference could also be due to a dependence of seeing upon
wavelength.

\begin{figure}
\epsfxsize=0.47\textwidth \epsfbox{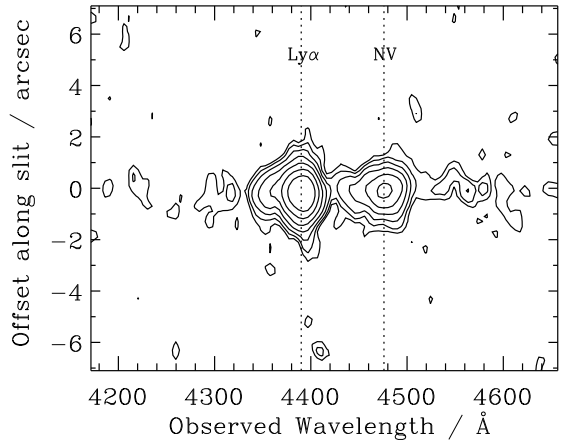}
{\caption[junk]{\label{fig:lyaln225}Two-dimensional spectrum of the
Ly$\alpha$ and N\,V emission lines of N2\_25. The vertical dotted
lines correspond to the expected positions of these lines for a
redshift $z=2.610$. Both emission lines are quite asymmetric with a
broader wing on the blueward side.}}
\end{figure}

Fig.\ \ref{fig:lyaln225} shows the two-dimensional spectrum of the
Ly$\alpha$ and N\,{\small V} region. It is clear that both lines are
asymmetric with a broader wing on the blueward side. These blue wings
appear as residuals in the gaussian fitting process which gave the
FWHM values in Table \ref{tab:linesn225}. The velocity offsets of
these residuals from the line peaks at $z=2.610$ are $\approx -2400$
km s$^{-1}$ for both Ly$\alpha$ and N\,{\small V}. These components
could be associated with an outflowing line-emitting wind or an
unobscured line-of-sight to part of a higher velocity region of the
BLR.

The central wavelength of the gaussian fit to the H$\alpha$ line gives
a redshift of $z=2.617 \pm 0.001$, significantly greater than the
redshift measured from the other lines. The fact that the [O\,{\small
I}] line in the near-IR spectrum has the same redshift as the UV lines
in the optical spectrum shows that this is not due to wavelength
calibration differences.  The offset of H$\alpha$ from the other lines
is $+550$ km s$^{-1}$. It is possible that contamination of the broad
H$\alpha$ line by the narrow [N\,{\small II}] lines could be affecting
the position of the line centre. To investigate this possibility we
have fit the spectrum with a broad H$\alpha$ line of variable width
and flux and the two [N\,{\small II}] lines with fixed widths
equivalent to that observed for narrow [O\,{\small I}] (the fluxes of
the lines are variable but the ratio of the [N\,{\small II}] lines are
fixed such that $\lambda 6583 / \lambda 6548 = 3$). It is not possible
to get a good fit to the observed spectrum by fixing the redshifts of
all 3 lines at $z=2.610$. The only way to obtain a good fit is to
allow the redshift of the broad H$\alpha$ line to float. In this case
the best fit is obtained for a negligible contribution from the
[N\,{\small II}] lines and the H$\alpha$ line centre has the offset of
$+550$ km s$^{-1}$ as in Table \ref{tab:linesn225}. A $\approx +500$
km s$^{-1}$ shift in the Balmer lines compared to the narrow lines has
been observed for a sample of luminous quasars at $z \approx 2$,
although its cause remains unclear (McIntosh et al. 1999a). However,
the high-ionization broad UV lines are normally blue-shifted from the
narrow line redshift, but this is not observed for N2\_25. Finally we
note that there is a bump in the $K$-band spectrum at 2.3530 $\umu$m
which is possibly the H$\alpha$ counterpart of the Ly$\alpha$ and
N\,{\small V} residuals mentioned previously. The velocity offset from
$z=2.610$ of this H$\alpha$ component is $\approx -2100$ km~s$^{-1}$,
very close to the offset of the broad wing components in the
Ly$\alpha$ and N\,{\small V} lines.

\begin{figure}
\hspace{0.3cm}
\includegraphics[width=6cm,angle=270]{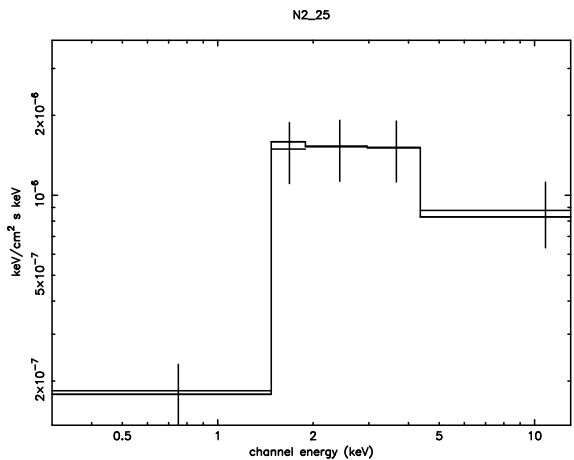}
\caption{\label{fig:xspec25}{\chandra ACIS-I spectrum of N2\_25. The error bars show the
binned data and the solid line is the best-fit model for an absorbed
power-law at $z=2.61$ with an intrinsic photon index of
$\Gamma=1.7$. The absorbing column density is $N_{\rm H}= (2.9 \pm
0.7) \times 10^{23} \, {\rm cm}^{-2}$.}}
\end{figure}

\subsection{The X-ray spectrum}
\label{n225xray}

N2\_25 has a very hard X-ray spectrum as indicated by its hardness
ratio of $0.26 \pm 0.11$ (Table \ref{tab:sample} -- see Manners et
al. 2002 for the hardness ratio definition and spread in values for the
whole survey). Spectral analysis of the \chandra data on this source
was performed using the {\footnotesize XSPEC} package. Due to the
relatively low number of net counts detected (85) it was not possible
to simultaneously constrain the power-law index and the absorbing
column. Therefore we set the photon index to be typical of an
unabsorbed quasar ($\Gamma=1.7$) and allowed the gas absorption column
$N_{\rm H}$ (at redshift $z=2.61$) to be fitted as a free
parameter. The best fit value is $N_{\rm H}= (2.9 \pm 0.7) \times
10^{23} \, {\rm cm}^{-2}$ (shown in Fig.\ \ref{fig:xspec25}).  For
comparison, using $\Gamma=2.0$ and $\Gamma=1.5$ resulted in column
densities of $N_{\rm H}= (3.8 \pm 0.8) \, {\rm and} \, (2.4 \pm 0.6)
\times 10^{23} \, {\rm cm}^{-2}$, respectively. Therefore for any
reasonable power-law continuum slope there is very heavy absorption by
a column with $\approx 3 \times 10^{23} \, {\rm cm}^{-2}$. The
observed hard X-ray (2-10 keV) rest-frame luminosity of N2\_25 is $1.6
\times 10^{44}$ erg s$^{-1}$, but correcting for the hard X-ray
absorption by this column density gives the corresponding de-absorbed
$2-10$ keV luminosity of $L_{\rm X}=4.7 \times 10^{44}$ erg s$^{-1}$.

\subsection{Gas and dust obscuration}

These observations of N2\_25 allow us to estimate the nuclear
obscuration from three different components: the X-ray spectrum, the
rest-frame optical-UV spectrum and the broad line region. Assuming
that the intrinsic X-ray spectrum is heavily absorbed by a column
density of $3 \times 10^{23} \, {\rm cm}^{-2}$, the corresponding
nuclear reddening would be $A_V \approx 150$, assuming a galactic
gas-to-dust ratio. Comparison of the optical spectrum with that of a
reddened typical quasar spectrum (Francis et al. 1991) shows that the
optical reddening is much lower with $A_V=1.4$ (or equivalently,
$E_{B-V}=0.45$ assuming a galactic extinction law:
$A_V/E_{B-V}=3.1$). The flux ratio of Ly$\alpha$ / H$\alpha =3$. For
luminous quasars, McIntosh et al. (1999b) find that this ratio is
typically $\approx 10$. Therefore there does not seem to be evidence
for a very large amount of reddening of the broad line region. The
amount of reddening required to change the observed value of this
ratio from $3$ to the typical value of $10$ is only $A_V=0.5$.

Reconciling the much lower reddening of the optical light (both
continuum and from the BLR) with the very high X-ray absorption leads
to two possibilities: either the absorbing medium has a very low
reddening to absorption ratio (about 1\% of the standard galactic
value), or the optical radiation does not pass through the bulk of the
medium responsible for the X-ray absorption. In the latter case, the
optical emission we observe would be scattered into our line-of-sight
and likely to be just a small fraction of the total
luminosity. Therefore a test of the scattering hypothesis is to see
whether the de-reddened optical luminosity (in lines or continuum) is
consistent with that expected given the absorption-corrected hard
X-ray luminosity.

It is well known that broad emission line luminosities, such as that
of H$\alpha$, correlate well with hard X-ray luminosity, due to the
direct link between these properties of photo-ionization. The ratio of
$L_{\rm X (2-10)} /L_{\rm broad H\alpha}$ can be used as an indicator
of the optical absorption to the BLR in Seyfert 2s and ULIRGs because
hard X-rays suffer much less absorption than the optical (Ward et
al. 1988). Imanishi \& Ueno (1999) give the best-fit linear
relationship between these quantities for broad-line AGN as $L_{\rm
X}/L_{\rm broad H\alpha}=18$ with a small rms scatter of a factor of
$\approx 2$. For N2\_25 we have determined the broad H$\alpha$
luminosity to be $3.2 \times 10^{43}$ erg s$^{-1}$ which would
increase up to $9.0 \times 10^{43}$ erg s$^{-1}$ if we take account of
the reddening of the optical continuum ($A_V=1.4$). Using the
de-absorbed hard X-ray luminosity of $4.7 \times 10^{44}$ erg s$^{-1}$
gives a ratio of $L_{\rm X}/L_{\rm broad H\alpha}=5$ (or 15 if we do
not take account of BLR reddening). This implies that we are seeing
effectively all the H$\alpha$ emission, ruling out a scattering origin
for this line.

Quasars tend to follow a well-defined correlation between the X-ray
flux and optical magnitude (e.g. Schmidt et al. 1998) and therefore we
can apply a similar test to that in the previous paragraph using the
optical continuum.  The observed ratio of hard X-ray flux to optical
(observed $R$-band) flux of N2\_25 is $10 \times$ greater than for
typical quasars (Willott et al. 2001a; Gonz\'alez-Solares et
al. 2002). Correcting for the hard X-ray absorption of a factor of 3
gives a ratio of X-ray to optical flux $30 \times$ greater than
typical, i.e. 3.7 magnitudes. The observed optical extinction of
$A_V=1.4$ corresponds to 3.3 magnitudes of extinction at the observed
$R$-band for a galactic extinction law. Therefore, after corrections,
the intrinsic ratio of hard X-ray to optical fluxes of N2\_25 is
typical of quasars. This result is consistent with that found for the
ratio of X-ray and broad emission line fluxes.

The results above suggest that once we take account of a small amount
of reddening ($A_V \approx 1$) and the much greater X-ray absorption,
we find the intrinsic hard X-ray, optical continuum and broad-line
luminosity ratios are all consistent with those found in normal
quasars.  Hence it does not appear that the optical light we see is
scattered and we deduce that both the optical and hard X-ray radiation
are passing through the same material. 

In recent years, there have been several reports of AGN which tend to
show much more absorption in X-rays than in their optical properties
(Simpson 1998; Akiyama et al. 2000; Risaliti et al. 2001; Maolini et
al. 2001a; Comastri et al. 2001).  Possible explanations of this
effect are discussed in Maiolino et al. (2001a,b). The most likely
reasons for the higher gas absorption are that either the gas-to-dust
ratio is much higher in the circumnuclear regions of AGN than the
standard galactic ratio or the dust grain composition is
different. The implication of such a scenario is that the AGN
classification into Types 1 and 2 could be decoupled in the optical
and X-rays, such that there exist a large population of sources with
properties similar to N2\_25. The existence of large numbers of
optically-bright quasars with hard X-ray spectra would lessen the
number of optically-obscured QSOs necessary in models which fit the
X-ray background. In future work we plan to investigate the fraction
of such objects in the EDXS.

\subsection{Interpretation}

As shown in Section \ref{n225spec}, the FWHM of the broad emission
lines in N2\_25 are very narrow ($\approx 2000{\rm ~km~ s}^{-1}$) for
such an intrinsically luminous quasar. This low velocity width could
arise from three different possibilities: (i) a small black hole mass
$M_{\rm BH}$, since assuming Keplerian motion the FWHM $\propto M_{\rm
BH}^{0.5}$; (ii) a pole-on line-of-sight, since the BLR gas is likely
to be in a flattened disk configuration along the torus axis and
therefore when viewed pole-on there is a reduced radial velocity
component (Wills \& Browne 1986); (iii) there is
substantial obscuration of the BLR and we are only viewing some
regions with low velocity dispersion.

In case (iii) we would expect the emission line profiles to be poorly
fit by a single gaussian. Indeed, there is some evidence of this from
the blueshifted (by $2500 {\rm ~km~ s}^{-1}$) components observed in
the wings of Ly$\alpha$, N\,{\small V} and possibly
H$\alpha$. However, this blueshifted emission could equally be due to
an outflow. There is some observational support for separating the
broad line region of quasars into an intermediate line region (ILR)
and very broad line region (VBLR) (Brotherton 1996). The ILR is
proposed to be further away from the active nucleus and has lower
velocity dispersion ($\sim 2000 {\rm ~km~ s}^{-1}$) and stronger low
ionization lines. In this case our observed small velocity widths
would be due to total obscuration of the VBLR. However, the observed
emission line spectrum of N2\_25 is characterized by strong highly
ionized species such as N\,{\small V} and O\,{\small VI} which are
expected to be relatively weak in the ILR. As mentioned in the
previous section, the absorption-corrected ratios of X-ray and broad
line luminosities are typical and therefore suggest that the majority
of the broad line emission is not totally obscured.

Discounting the VBLR obscuration hypothesis leaves us with the
possibilities of either a relatively small black hole mass or an
orientation effect. Considering the first of these, one can calculate
the black hole mass assuming Keplerian motion of the BLR clouds and
that N2\_25 has no special orientation with respect to us. McLure \&
Jarvis (2002) give the relationship between black hole mass, optical
luminosity and BLR FWHM as $M_{\rm BH} / {\rm M_{\sun}} =4.74 (
\lambda L_{\rm 5100} /10^{37} {\rm W})^{0.61} ({\rm FWHM / km~
s}^{-1})^2 $. For N2\_25 with ${\rm FWHM}=2000 {\rm ~km~ s}^{-1}$ we
find $M_{\rm BH}=1.5 \times 10^{8} {\rm M_{\sun}}$. Given the
absorption-corrected absolute magnitude of $M_B=-24.9$, we determine
its bolometric luminosity to be $3 \times 10^{39}$ W (assuming a
bolometric correction, $L_{\rm bol}=10 \lambda L_{\rm 5100}$ as in
Wandel, Peterson \& Malkan 1999). The Eddington limit for a black hole
mass of $M_{\rm BH}=1.5 \times 10^{8} {\rm M_{\sun}}$ is $2 \times
10^{39}$ W, therefore N2\_25 would be accreting at approximately its
Eddington limit, higher than most low-redshift quasars which are
typically accreting at $\approx 10$\% of the Eddington limit
(e.g. Kaspi et al. 2000; McLure \& Dunlop 2001).

N2\_25 shares several properties with the class of objects known as
narrow-line Seyfert 1s (NLSy1s) such as narrow emission lines and very
high metallicity (as indicated by the strength of N\,{\small V}). A
major difference between N2\_25 and NLSy1s is that the NLSy1s tend to
have a soft X-ray excess and steep hard X-ray spectra ($\Gamma \approx
2.2$; Brandt, Mathur \& Elvis 1997). However, any soft X-ray excess
would be redshifted out of the \chandra energy range and the spectrum
of N2\_25 can be fit with an intrinsically steep hard X-ray spectrum
with a slightly higher absorbing column than that derived in Section
\ref{n225xray}. The relatively small black hole for such a luminous
quasar could indicate that the black hole is still in a growing phase,
since there is evidently a plentiful fuel supply, as has been proposed
for NLSy1s (Mathur 2000).

\begin{figure*} 
\vspace{0.2cm} 
\hspace{-0.7cm}
\epsfxsize=0.97\textwidth 
\epsfbox{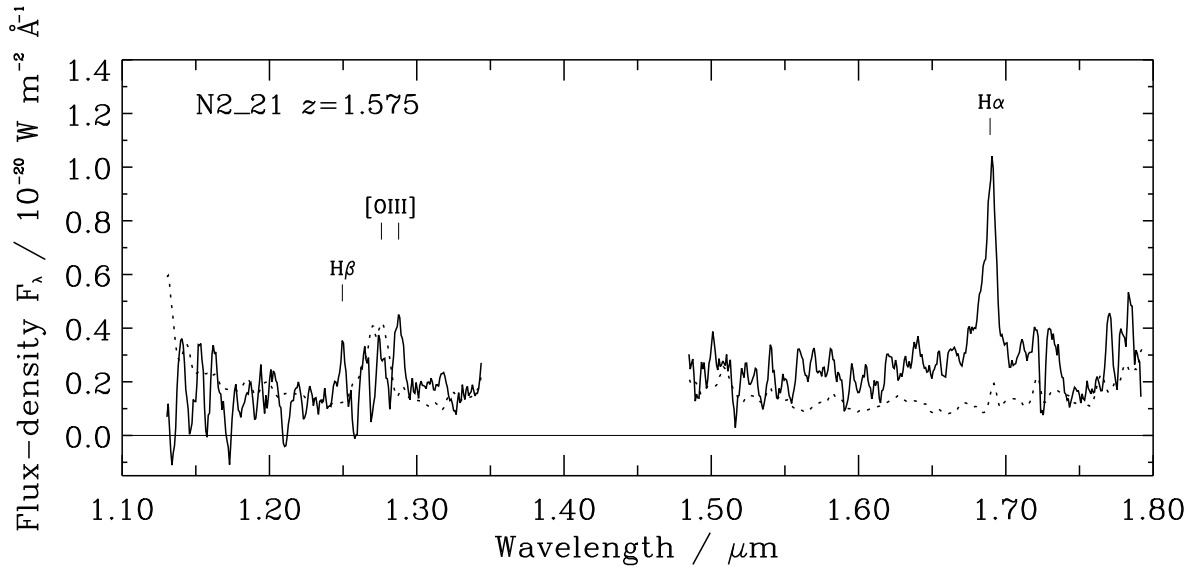} 
\vspace{-0.5cm}
{\caption[junk]{\label{fig:specn221} OHS spectrum of the radio galaxy
N2\_21 with emission lines labelled. The rms noise in the spectrum is
shown as a dotted line. Note that the [O\,{\small III}] 5007 \AA\ and
H$\beta$ lines lie in regions of low noise (either side of the
$1.27\umu$m O$_2$ sky emission feature) and hence both are securely
detected.}}
\end{figure*}

As suggested previously, the small FWHM of N2\_25 may actually be
indicative of an orientation effect, rather than related to the mass
of the black hole and the fraction of the Eddington luminosity. If we
are observing N2\_25 within a fairly small angle to the torus axis,
then the derived black hole mass could be a factor of $\approx 5
\times$ greater (e.g McLure \& Dunlop 2001), thereby reducing the
percentage of the Eddington luminosity at which the quasar is
accreting to $30$\%, similar to other quasars. However, for distant
radio-quiet quasars there is no orientation indicator as exists for
radio-loud objects, so the orientation remains unknown. Under the
unification scenario, the dusty torus provides the bulk of the X-ray
and optical obscuration. For radio-loud objects it has been found that
quasars observed close to the torus opening-angle tend to be redder,
due to dust reddening by material near the edges of the torus (Baker
1997). However, Becker et al. (2000) find that about a third of
radio-loud broad absorption line quasars have flat radio spectra,
signifying that these obscured objects are observed pole-on. It would
be interesting therefore if N2\_25 is actually observed close to
pole-on and contains a very high column density absorber.

N2\_25 is detected in our deep 1.4-GHz VLA map (Ivison et al. 2002)
with a peak flux-density of $133 \pm 9 \umu$Jy. The source is
spatially unresolved at 1.5 arcsec resolution, implying an angular
size $\ltsimeq 1$ arcsec.  The radio luminosity (evaluated at 5-GHz to
minimize $k$-corrections) is $L_{\rm 5-GHz}=2.1 \times 10^{24} {\rm W
Hz}^{-1}$. This is below the usual dividing line between radio-loud
and radio-quiet quasars of $L_{\rm 5-GHz}=10^{25} {\rm W Hz}^{-1}$
(e.g. Kellermann et al. 1994) and this quasar is classified as
radio-quiet.

N2\_25 is marginally detected in the mid-IR with ISOCAM, with a $15
\umu$m flux of $0.49 \pm 0.18$ mJy. The ratio of
mid-IR to hard X-ray fluxes for this source is similar to that of 
AGN in the Lockman Hole and Hubble Deep Field North (Fadda et
al. 2002; Alexander et al. 2002). At $z=2.61$, N2\_25 has the highest
known redshift of all the mid-IR sources within the 0.14 square
degrees of the ELAIS survey covered by the \chandra observations.

\section{N2\_21 - a dusty radio galaxy}

The X-ray source N2\_21 coincides with the position of a bright radio
source ($S_{1.4}=119$ mJy; Ciliegi et al. 1999). The optical
counterpart to the X-ray and radio source is a faint galaxy
($R=25.02$). The $K$-band image also shows a resolved counterpart with
$K=18.60$ giving an extremely red colour of $R-K=6.4$.

\subsection{Spectroscopic data}
\label{n221spec}

The OHS spectrum of N2\_21 is shown in Fig.\ \ref{fig:specn221}.  A
strong emission line of H$\alpha$ is detected at $1.690 \,
\umu$m. Weak emission lines at the expected positions of H$\beta$ and
[O\,{\small III}] are also detected. Table \ref{tab:linesn221} gives
the emission line data.  We adopt a redshift of $z=1.575$ on the basis
of the H$\alpha$ line position, since this is detected with the
highest signal-to-noise.  The widths of the emission lines are all
consistent with being unresolved at the spectral resolution of 80
\AA\, ($1500$ km\,s$^{-1}$).  This indicates these lines are narrow
and typical of those of high-redshift radio galaxies. The H$\alpha$
emission line is spatially unresolved on the 2-dimensional
spectrum. The Balmer decrement ratio of the fluxes of the H$\alpha$
and H$\beta$ lines is $\approx 9$ (note that this assumes the
contribution of the nearby [N\,{\small II}] lines to the H$\alpha$
flux is negligible). This is much larger than the typical case B
recombination value of 3, perhaps suggesting that there is some
reddening [$E(B-V) \approx 1$] along our line-of-sight to the narrow
line region. However, the low signal-to-noise of the H$\beta$ line,
the possibility of contamination of H$\alpha$ by [N\,{\small II}] and
the uncertainty of the intrinsic line ratio prevents a firm
statistical limit on the reddening being derived. The optical spectrum
of N2\_21 showed neither continuum nor emission lines. This is
consistent with the lack of strong emission lines in the optical
wavelength range for this redshift and does not constrain the
reddening to the narrow line region.

\begin{table}
\footnotesize
\begin{center}
\begin{tabular}{lcccc}
\hline
\mc{1}{l}{Emission line} &\mc{1}{c}{$\lambda_{\rm obs}$} &\mc{1}{c}{$z_{\rm em}$} &\mc{1}{c}{FWHM}&\mc{1}{c}{Flux/$10^{-19}$ }  \\
\mc{1}{c}{ } &\mc{1}{c}{(\AA)} &\mc{1}{c}{ } &\mc{1}{c}{(\AA)} &\mc{1}{c}{(W m$^{-2}$)} \\
\hline
H$\beta\,          \lambda 4861$ &   12496 $\pm$ 8 & 1.571  & $64 \pm 14$  &  0.9 $\pm$ 0.4 \\
${\rm [O\,III]}\,  \lambda 5007$ &   12878 $\pm$ 8 & 1.572  & $68 \pm 10$  &  2.0 $\pm$ 0.5 \\
H$\alpha\,         \lambda 6563$ &   16899 $\pm$ 9 & 1.575  & $87 \pm 10$  &  8.0 $\pm$ 1.2 \\
\hline
\end{tabular}
{\caption[junk]{\label{tab:linesn221} Emission line data for N2\_21
measured from the OHS spectrum.  All line widths are consistent with
being unresolved at the instrumental resolution.  The spectrograph
slit was oriented at a position angle $5^{\circ}$ east of north for
these observations.}} 
\normalsize
\end{center}
\end{table}

Assuming that N2\_21 contains an obscured broad line region as
expected in unified schemes, we can obtain a limit on the differential
reddening between the broad and narrow line regions. From the lack of
a broad base to the narrow H$\alpha$ line in Fig.\ \ref{fig:specn221},
the observed ratio of broad to narrow line fluxes is $<1$. Assuming an
intrinsic ratio of 40 for unobscured radio-loud quasars (Jackson \&
Eracleous 1995), this gives a lower limit to the differential
reddening of $A_V>5$. Combined with the estimated extinction to the
narrow line region of $A_V \approx 3$, gives that the extinction along
our line-of-sight to the broad line region is $A_V>8$.

\subsection{Radio imaging}
\label{n221radio}

The radio source coincident with N2\_21 is unresolved in our deep
1.4-GHz VLA map which has 1.4 arcsec resolution (Ivison et al. 2002).
The radio spectrum is steep between 326-MHz and 1.4-GHz with
$\alpha=0.87$. In order to determine the radio structure, we obtained
higher resolution data with the VLA at 8.4-GHz. The source was
observed for two 5 minute snapshots, using $2 \times 50$ MHz
bandwidths centred at 8.4351 and 8.4851 GHz. The data was flux
calibrated using 3C~286 and phase calibrated using the VLA calibrator
1640+397.  The data was self calibrated (for phase only), and imaged
using the {\footnotesize AIPS} task {\footnotesize IMAGR} with a
{\footnotesize ROBUST} parameter = 0, i.e. a compromise between
natural and uniform weighting.  The resulting synthesized beam is $0.23 \times
0.21$ arcsec at a position angle of $-28.8^{\circ}$. The rms noise in
the final map is 60 $\umu$Jy/beam.

The optical and near-infrared images were aligned with the radio
reference frame by comparing the positions of 5 radio sources on the
1.4-GHz image within a 2 arcmin radius around N2\_21 with their
optical counterparts. These sources (2 quasars and 3 galaxies) all
have fairly compact radio and optical structures. The rms in the
offsets between the radio and optical positions is only 0.12 arcsec in
RA and 0.10 arcsec in DEC. 

\begin{figure} 
\vspace{0.2cm} 
\hspace{-0.7cm}
\epsfxsize=0.47\textwidth 
\epsfbox{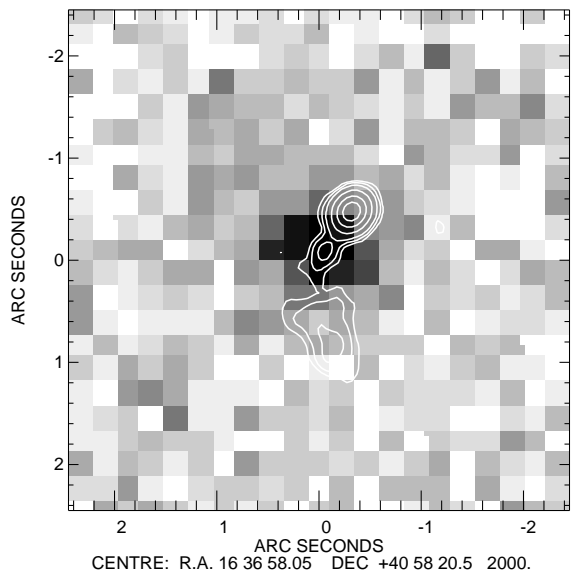} 
\vspace{-0.5cm}
{\caption[junk]{\label{fig:radn221} 8.4-GHz VLA image of the radio
galaxy N2\_21 (contours) overlaid on the INGRID $K$-band image
(greyscale). Radio contours are plotted at 3,6,12,36,96,192 $\times
\sigma$ where $\sigma=60 \umu$Jy. The radio structure shows a weak
core which is coincident with the near-infrared counterpart and
asymmetric lobes/hotspots to the north-west and south. The radio
observations are described in Sec.\ \ref{n221radio}.}}
\end{figure}

The 8.4-GHz map of N2\_21 is shown in Fig.\ \ref{fig:radn221} along
with the registered $K$-band image. There are three distinct radio
components which resemble the core and lobes/hotspots of powerful
radio galaxies. The weak core is unresolved and has a flux-density of 0.9
mJy. Its position of 16:36:58.05 +40:58:20.6 is identical to that of
the centroid of the $K$-band emission, which is as expected for core
emission from the bases of the jets close to the supermassive black
hole at the centre of the galaxy. The northern hotspot lies 0.40 arcsec
away and is significantly resolved with a deconvolved size of $0.1
\times 0.05$ arcsec along a position angle of 135 degrees east of
north, i.e. pointing towards the core. The fainter, southern lobe is
1.1 arcsec from the core and much more extended. The northern hotspot
dominates the radio flux with an integrated flux-density of 19.7 mJy out of
the total $26.9 \pm 0.5$ mJy of the whole source. It also dominates
the flux at 1.4-GHz given the 1.4-GHz position of 16:36:58.03
+40:58:20.8. The spectral index of N2\_21 between 1.4-GHz and 8.4-GHz
is steep with $\alpha=0.83$, consistent with the same slope as at
lower frequencies.

The core to lobe flux ratio is similar to other radio galaxies of this
redshift and luminosity (Blundell et al. in prep.). This, combined
with the spectroscopic observations discussed in the previous
section, lead us to conclude that N2\_21 appears to be a fairly
typical radio galaxy and does not have the properties of a quasar
(such as bright radio core and broad emission lines). One unusual
aspect of the radio structure is that the lobes/hotspots are quite
asymmetric in terms of position angle, distance from the core and
flux-density. These asymmetries are not due to the beaming of the
hotspot emission, since this is not thought to be a major factor for
radio galaxies (Dennett-Thorpe et al. 1999). The observed asymmetries
would tend to suggest that the radio source is expanding into an
inhomogeneous medium (Barthel \& Miley 1988). The side of the
brighter, shorter hotspot would then have a much denser ambient medium
which confines the hotspot advance and enhances the radiative
efficiency. McCarthy, van Breugel \& Kapahi (1991) also find that the
closer hotspot tends to have a higher optical emission line
luminosity, consistent with the idea of a denser medium. It is
interesting to note that the faint $R$-band emission in N2\_21, which
samples the rest-frame UV, is more extended than the $K$-band emission
and centred closer to the north hotspot than the core.

\subsection{The X-ray spectrum and plausible emission mechanisms} 
\label{n221xray}

\begin{figure}
\hspace{0.3cm}
\includegraphics[width=6cm,angle=270]{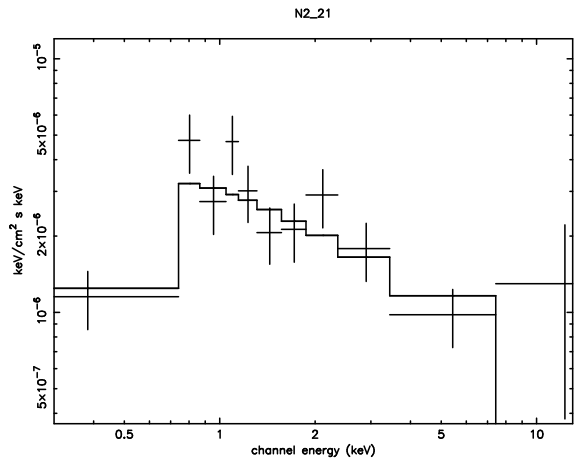}
\caption{{\label{fig:xspec21}ACIS-I spectrum of
N2\_21. The error bars show the binned data and the solid line is the
best-fit model for an absorbed power-law at $z=1.57$ with an intrinsic
photon index of $\Gamma=1.7$. The absorbing column density is $N_{\rm
H}= (5.4 \pm 3.1) \times 10^{21} \, {\rm cm}^{-2}$, i.e. the spectrum
is consistent with negligible absorption.}}
\end{figure}

The X-ray spectrum of N2\_21 is fairly soft with $HR=-0.42 \pm 0.07$
(a total of 175 net counts were detected).  The best-fit power-law
spectrum in {\footnotesize XSPEC} gives $\Gamma=1.80 \pm 0.22$ and
$N_{\rm H}= (7.0 \pm 5.1) \times 10^{21} \, {\rm cm}^{-2}$. The hard
($2-10$ keV) X-ray luminosity of N2\_21 is $L_{\rm X}=2 \times
10^{44}$ erg s$^{-1}$. Fig.\ \ref{fig:xspec21} shows the binned data
with a model where the photon index is fixed to that of a typical
quasar ($\Gamma=1.7$) giving $N_{\rm H}= (5.4 \pm 3.1) \times 10^{21}
\, {\rm cm}^{-2}$. The X-ray spectrum is similar to that of a typical
quasar with only a very small amount of absorption ($\approx 50
\times$ lower than for N2\_25). We have also fitted the X-ray spectrum of
N2\_21 in {\footnotesize XSPEC} with a thermal Raymond-Smith model and find
that viable fits are only obtained for very high temperatures (the
best fit is for T=18 keV with reduced $\chi^2=1.0$ and T=10 keV has
reduced $\chi^2=1.3$). Much lower temperatures which are more common
in hot cluster gas are ruled out (e.g. a fit with temperature of 5 keV
has reduced $\chi^2=2.5$). This suggests a non-thermal origin for the
X-ray emission is more plausible.

The lack of X-ray absorption in N2\_21 is rather surprising given that
the fact we observe only narrow optical emission lines implies that
the broad line region is heavily obscured. \chandra observations of
the nuclei of other high-redshift radio galaxies show much harder
spectra with absorbing columns $ \gg 10^{22} \, {\rm cm}^{-2}$
(e.g. Fabian et al. 2001; Carilli et al. 2002). Assuming a galactic
gas-to-dust ratio, the reddening due to the gas column density in
N2\_21 is $E(B-V)\approx 1$ (note the X-ray spectrum is also formally
consistent with no absorption). This amount of reddening is consistent
with that from the narrow line Balmer decrement, suggesting it could
be due just to material in the extended host galaxy (see also Sec.\
\ref{n221host}). The limit on the reddening to the broad line region
derived in Section \ref{n221spec} implies a limit on the column
density of $N_{\rm H}> 1.5 \times 10^{22} \, {\rm cm}^{-2}$.  The fact
that this is considerably greater than the observed limit on the
column density raises the possibility that the bulk of the observed
X-ray emission is not from the nucleus, but from a more extended
region beyond the obscuring torus.

On the \chandra image, N2\_21 is consistent with being a point source
with a FWHM = 0.7 arcsec (equivalent to 6 kpc at $z=1.57$). We cannot
determine whether the centroid of the X-ray emission is closer to the
radio core or hotspot since their separation is only 0.4
arcsec. However, the X-ray position is not consistent with the southern
lobe/hotspot. The small scale of the X-ray emission argues against the
soft X-ray emission being due to large-scale thermal cluster gas which
would be expected to have a much greater extent of $\sim 100$
kpc. However, it is possible that the soft X-ray emission is related
to the bright, compact radio emission from the northern hotspot. We
can rule out the possibility that the X-rays are due to an
extrapolation of the synchrotron radiation since the radio to X-ray
spectrum would need to have a slope of $\alpha=1.3$ to give the
required X-ray flux and the X-ray spectral index of $0.8$ is flatter
than this. Further, we can rule out Thomson scattering of nuclear
X-ray emission since scattering fractions are generally rather low
(e.g. 1\% for the central few kpc of Cygnus A -- Young et al. 2002)
and hence the intrinsic hard X-ray luminosity would have to be
$>10^{46}$ erg s$^{-1}$. However, given the [O\,{\small III}] emission
line luminosity of $3 \times 10^{42}$ erg s$^{-1}$, the correlations
between [O\,{\small III}] and hard X-ray luminosities for both radio
galaxies and Seyferts predict that the intrinsic hard X-ray luminosity
is only $\sim 10^{44}$ erg s$^{-1}$ (Sambruna, Eracleous \& Mushotzky
1999; Mulchaey et al. 1994).

Recent \chandra imaging of $z \approx 2$ radio-loud AGN has revealed
that extended soft X-ray emission on scales and axes similar to the
radio lobes are a common occurrence (e.g. 3C~294 -- Fabian et al. 2001;
PKS~1138-262 -- Carilli et al. 2002; 3C~9 - Fabian et al. 2002).
Unfortunately, the low numbers of counts detected in each case limit
the constraints on the spectral shapes of this extended emission,
except that it is fairly soft in all cases.  The most likely emission
mechanisms invoked by the authors are different for all three
sources. For 3C~294 it is thermal cluster gas. For PKS~1138-262 it is
gas which has been shocked by the expanding radio source. For 3C~9, it
is claimed that the double-sided jet is scattering nuclear
photons.

The spectrum of N2\_21 cannot be fit well with a thermal model, but
the reduced $\chi^2=1.3$ for hot ($T=10$ keV) gas is not strictly
ruled out (note also that a lower temperature gas could also fit the
data if it is absorbed by a high column density). Therefore we keep
the option open that the X-ray emission could be due to gas shocked by
the expanding radio source (c.f. Carilli et al. 2002) until we have
more sensitive, higher spectral resolution X-ray data. The compactness
of the X-ray emission indicates that to produce such a high X-ray
luminosity the gas would need to be very dense, $\sim 10^8$
m$^{-3}$. The density of dark matter (and gas) within recently
collapsed structures scales with redshift as $(1+z)^{3}$. The
arm-length and flux asymmetries indicate a difference in the densities
of the two sides $\sim 10$ assuming models for radio source evolution
and emission (e.g. Willott et al. 1999). Strong shocks from the
advancing radio source could provide a further increase of a factor of
$\approx 4$ in density.  Together these factors give a density which
is $\sim 1000$ higher than is found in local clusters. Young et
al. (2002) estimate the central electron density in the cluster
surrounding Cygnus A is $\sim 10^5$ m$^{-3}$. Therefore it is
plausible that the high density required by the hot gas model to
explain the X-ray emission from N2\_21 could be viable. One
consequence of this high density is that the cooling time would be
very short, but this is not a problem as the small angular size of the
radio source indicates its jet-triggering event was quite recent and
energy is constantly being pumped into the environment by the
expanding radio source.

We next consider inverse--Compton (IC) scattering off the relativistic
electrons in the radio lobes as the X-ray production mechanism for
N2\_21. If the relativistic electrons and seed photons are co-spatial
in a uniform magnetic field then the IC scattering luminosity is
$L_{\rm IC}=L_{\rm synch} \times (U_{\rm ph} / U_{\rm B})$ where
$L_{\rm synch}$ is the integrated synchrotron luminosity, $U_{\rm ph}$
the energy density in the seed photons and $U_{\rm B}$ the energy
density in the magnetic field (e.g. Moran, Lehnert \& Helfand
1999). In such a compact radio source, the energy density of the IR
quasar photons is likely to dominate over that of either the cosmic
microwave background or the synchrotron emission itself (Brunetti et
al. 2001). Since $L_{\rm X} \gg L_{\rm synch}$ for N2\_21, then the IC
mechanism can only produce a high enough luminosity if $U_{\rm ph} \gg
U_{\rm B}$. This will only be true for a magnetic field strength much
lower than the equipartition value. Blundell \& Rawlings (2000) argue
that such deviations from equipartition will only occur in large, aged
radio sources and not in compact, young sources. Therefore we find
that IC scattering of the nuclear photons cannot generate all the
observed X-ray emission.

In conclusion, the emission mechanism for the X-rays from N2\_21
remains a mystery. The compact nature of the radio and X-ray emission
precludes differentiating between nuclear emission or emission on a
size scale comparable to the radio source. No viable mechanisms
related to the radio source seem to be able to fit the spectrum and
extent of the soft emission from this object. Perhaps we cannot
exclude the possibility that the X-rays come from the nucleus and are
not heavily obscured, despite the obvious obscuration in the optical
and the weakness of the radio core. Whereas N2\_25 showed evidence for
a very high gas-to-dust ratio in its absorbing material, perhaps the
absorber in N2\_21 has a very low gas-to-dust ratio such that we
observe most of the soft X-rays coming from the nucleus. As noted
above, the observed X-ray luminosity of N2\_21 is consistent with that
expected given its narrow emission line luminosity.

\subsection{The evolutionary state of the host galaxy}
\label{n221host}

The sensitive near-infrared spectrum and acquisition of a redshift for
N2\_21 allows us to investigate the properties of the stellar
continuum in this galaxy. Willott et al. (2001b) showed that about half
of all radio galaxies at $1<z<2.5$ have extremely red colours
($R-K>5.5$). However, for these objects, spectroscopic redshifts were
not available and broad-band photometry alone did not allow a
distinction to be made between evolved stellar populations or dusty
starbursts. N2\_21 has properties such as redshift, $R-K$ colour and
radio luminosity similar to the extremely red radio galaxies in
Willott et al. (2001b).

\begin{figure} 
\vspace{0.5cm} 
\epsfxsize=0.48\textwidth 
\epsfbox{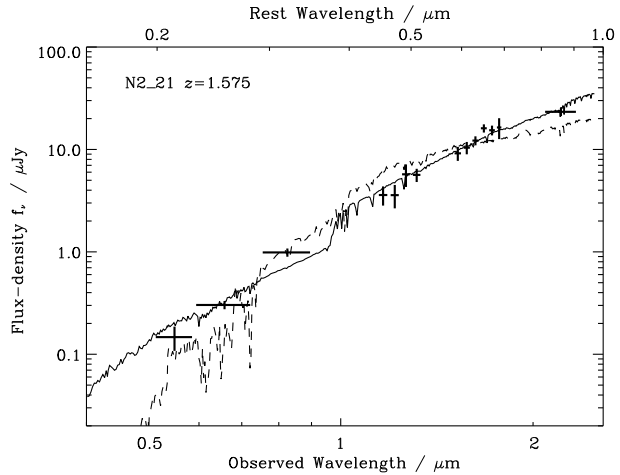} 
{\caption[junk]{\label{fig:sed221} Optical ($V,R,I$) and near-IR
($J,H,K$) photometry of the radio galaxy N2\_21. The $J$ and $H$ band
data are in 500 \AA\ bins from the OHS spectrum. Also shown are the
best fitting reddened (solid) and unreddened (dashed) template SEDs
for galaxies at this redshift. The reddened model has age\,=\,90\,Myr
and $A_V=2.6$ (Milky Way type dust) and the unreddened model has
age\,=\,1.7\,Gyr.  The poor fit of models without reddening indicates
a large amount of dust reddening in this galaxy and suggests it is
undergoing a starburst.}}

\end{figure}

The continuum spectral index in the $J$ and $H$ bands is $\alpha=3.5$.
This is considerably steeper than the $\alpha \approx 2$ of
unreddened, moderately old (a few Gyr) stellar populations at this
redshift and suggests that dust reddening is important.  Pozzetti \&
Manucci (2000) have proposed colour-colour criteria which can
distinguish between old and dusty EROs. They show that, because the
4000 \AA\ break in the spectrum falls between the $I$ and $J$ bands at
$z<2$, EROs with large $J-K$ colours compared to their $I-K$ colours
tend to be dusty starbursts and those with smaller $J-K$ colours
represent old galaxies. N2\_21 has $I=23.5,J=21.4,K=18.6$ giving an
$I-K$ colour of 4.9 and a $J-K$ colour of 2.8. This places it well
into the starburst regime (the division between old galaxies and dusty
starbursts for $I-K=4.9$ is $J-K=2.2$).

The images of N2\_21 at all optical and near-IR wavelengths show
resolved structure and hence we are confident that the observed light
is stellar in origin and not strongly affected by a point-source such
as a reddened quasar. Using the observed broad-band magnitudes in the
$V$, $R$, $I$ and $K$ bands and the calibrated OHS spectrum in the $J$ and
$H$ bands we have determined the rest-frame UV-optical spectrum of
N2\_21 (Fig.\ \ref{fig:sed221}). Emission lines were subtracted from
the OHS spectrum and fluxes in 500 \AA\, bins generated to reduce the
errors. The SED cannot be fit by a single power-law as might be
expected if scattered AGN emission dominates at all wavebands. 

A wide range of galaxy SEDs were fit to these data using the
{\footnotesize HYPERZ} routine (Bolzonella, Miralles \& Pell\'o
2000). The redshift was fixed to $z=1.57$ and a range of Bruzual \&
Charlot evolving SEDs were used.  The SEDs were subject to variable
amounts of reddening using both the Milky Way (MW) extinction curve
(Allen 1976)\footnotemark and that of starburst galaxies derived by
Calzetti et al. (2000). The maximum amount of reddening was determined
such that the de-reddened $K$-magnitude of N2\_21 would not be
brighter than the $K=17$ typically observed for the most luminous
radio galaxies at this redshift (Eales et al. 1997). This corresponds
to $A_V \leq 2.8$ for MW extinction or $A_V \leq 2.3$ for the Calzetti
law. The best fitting model (reduced $\chi^2 =1.5$) is for a 90 Myr
old starburst reddened by $A_V=2.6$ of MW dust. The best fitting model
with the Calzetti reddening law is for a fairly old (2.0 Gyr)
elliptical reddened by $A_V=2.2$ (reduced $\chi^2 =2.1$). Younger
starbursts with ages $\sim 10$ Myr can also fit the data, but require a higher
$A_V$ or a steeper extinction curve than the MW. Finally we performed
a fit without dust reddening and this gave a very poor fit with
reduced $\chi^2 =5.3$, confirming our expectation from the broad-band
colours and near-IR continuum slope that the starlight from N2\_21 is
strongly affected by dust. The best fit MW-reddened and unreddened
spectra are plotted on Fig.\ \ref{fig:sed221}.

\footnotetext{The strong absorption feature at 2175 \AA\ was removed
from the Milky Way extinction curve since it is generally not observed
in extragalactic objects (e.g. Pitman, Clayton \& Gordon 2000).}

The large amount of reddening necessary to fit the SED of N2\_21
indicates substantial quantities of dust distributed throughout the
galaxy. The existence of such a large amount of dust strongly suggests
that N2\_21 is undergoing an intense starburst. This galaxy is
contained within the area surveyed at 850$\umu$m with SCUBA in the 8
mJy survey of the ELAIS N2 region (Scott et al. 2002). N2\_21 is not
detected in the submillimetre to a $3 \sigma$ upper limit of 5.3 mJy
(Almaini et al. 2002). This non-detection is not at a sensitive enough
level to rule out a massive starburst in this galaxy, only giving
limits of star-formation rate $ \ltsimeq 1000 {\rm M}_{\sun} {\rm
yr}^{-1}$ and dust mass $M_{\rm d} \ltsimeq 3 \times 10^{8} {\rm
M}_{\sun}$ (assuming dust parameters as in Hughes, Dunlop \& Rawlings
1997). More sensitive observations are required to determine the
far-infrared output of N2\_21 and its star-formation rate, but it is
clear from our observations that it must be a very dusty galaxy. It is
interesting to note that the radio source hosts with the highest
submillimetre luminosities tend to contain the smallest (youngest)
radio sources (Willott et al. 2002a; Rawlings et al. 2002) and N2\_21
appears to be a further example of a very dusty young radio source.

\section{N2\_28 - an obscured AGN in a cluster of EROs?}

\begin{figure} 
\hspace{-0.5cm}
\epsfxsize=0.52\textwidth 
\epsfbox{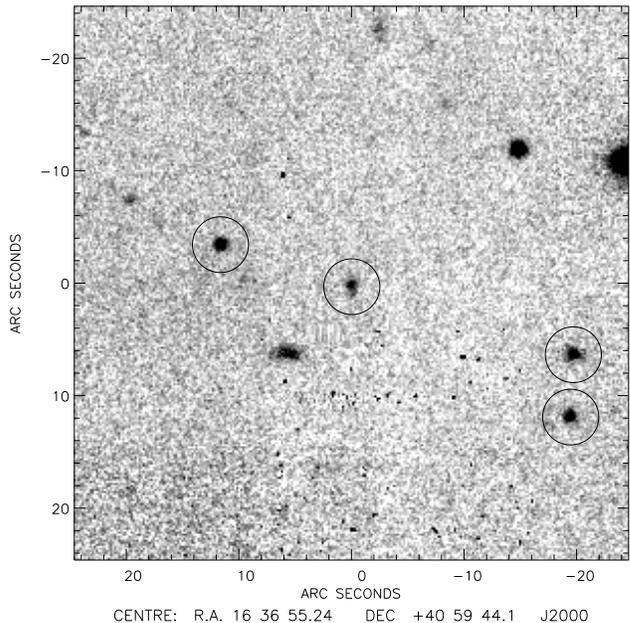} 
\vspace{-0.6cm}
{\caption[junk]{\label{fig:erocluster} UFTI $K$-band image of the
extremely red galaxy counterpart to N2\_28 (centre). Galaxies with
$R-K>6$ are circled. The extremely red galaxy to the east
of N2\_28 was put along the slit in the near-IR spectroscopy.}}

\end{figure}

The X-ray source N2\_28 is identified with an extremely red galaxy
which has $K=19.7$ and $R=25.8$. This galaxy is near the centre of a
group of 4 extremely red ($R-K>6$) galaxies (Fig.\
\ref{fig:erocluster}).  The other 3 extremely red galaxies are all a
bit brighter with $K \approx 19.2$. The surface density of objects
with $K<20$ and $R-K>6$ in this region is 30 times higher than is
typical (Thompson et al. 1999). This high overdensity strongly
suggests that at least some of these objects are at the same redshift
and physically associated. These galaxies are part of a cluster of
$\sim 10$ $R-K>5$ galaxies to the north-east of N2\_28 which is
discussed by Roche et al. (2002).

\subsection{Spectroscopic data}
\label{n228spec}

The slit of the OHS spectrograph was oriented at a position angle of
$72.4^{\circ}$ east of north for the observations of N2\_28 so that
the closest of the nearby EROs (object `r552' in Roche et al. 2002),
which is 12 arcsec north-east of the X-ray source, fell along the
slit. One-dimensional spectra of both sources were extracted from the
reduced 2-D spectra. Continuum was detected from both objects, but no
clear spectral features were found. The near-IR spectral indices of
both sources are $\alpha \approx 2$, much flatter than was found for
N2\_21, suggesting that the continua of these EROs are not so affected
by dust reddening. The lack of a 4000 \AA\ break further suggests that
these galaxies are not at very high redshift ($z>2$).

\subsection{The X-ray spectrum}
\label{n228xray}

N2\_28 is detected in both the hard and full band \chandra images.
The hardness ratio of $HR=-0.14 \pm 0.18$ is $2 \sigma$ harder than
the typical $HR=-0.5$ of unobscured quasars, indicating a substantial
absorbing column. We have fit the data in {\footnotesize XSPEC},
however with only 32 net counts detected from this source, our
constraints on the spectral shape are not very strong. An unobscured
power-law model has a best fit photon index of $\Gamma=0.8 \pm 0.3$.
To estimate the absorbing column if the source has an intrinsic
power-law similar to quasars ($\Gamma=1.7$), we also fit the spectrum
with a fixed slope and variable absorption at $z=1.1$. This gave a
best fit absorbing column of $N_{\rm H}= (4.5 \pm 2.6) \times 10^{22}
\, {\rm cm}^{-2}$. If the X-rays are from an obscured AGN, this
level of absorption is consistent with the optical and near-IR data
which shows a resolved host galaxy and no nuclear point-source. The
hard X-ray luminosity of N2\_28 is $L_{\rm X}=1.6 \times 10^{43}$ erg
s$^{-1}$ assuming it is at a redshift of 1.1 (see Sec.\
\ref{n228sed}), which would increase to $L_{\rm X}=2.2 \times 10^{43}$
erg s$^{-1}$, when correcting for the X-ray absorption. The hard
spectrum and high luminosity indicate the presence of an obscured AGN,
however this source is considerably less luminous than quasars and
more similar to Seyfert galaxies.

\subsection{Redshift constraints}
\label{n228sed}

We can use the information from the OHS spectrum and broad-band
photometry to determine the SEDs of N2\_28 and the nearby ERO placed
along the slit. These SEDs can then be compared to a range of galaxy
models to constrain the redshifts of these two galaxies. We have used
the {\footnotesize HYPERZ} package as described in Sec.\
\ref{n221host} to perform this analysis. Again the OHS fluxes in the
$J$ and $H$ bands were extracted in 500 \AA\ bins. Because the slope
of the near-IR spectra indicated that there is not a huge amount of
reddening of these sources, the maximum reddening used was
$A_V=1$. Both Calzetti and MW extinction laws were used, giving very
similar results, so for simplicity we will only quote the results
obtained for the Calzetti extinction law.

\begin{figure} 
\vspace{0.5cm} 
\epsfxsize=0.48\textwidth 
\epsfbox{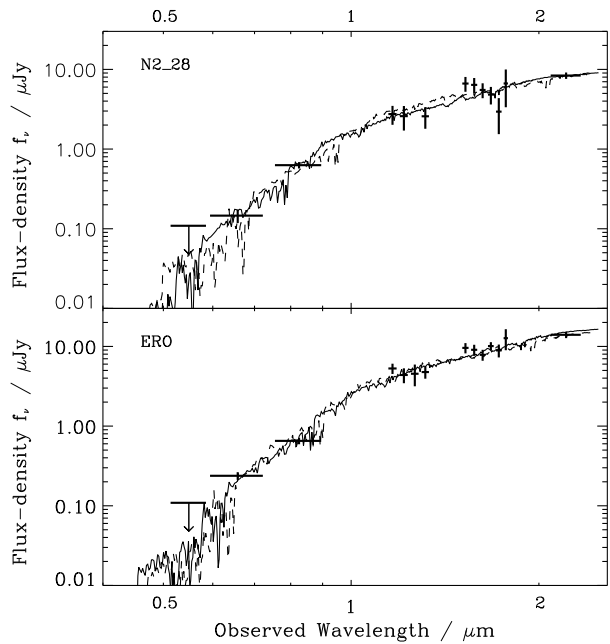} 
\vspace{0.6cm}
{\caption[junk]{\label{fig:sed228} Optical ($V,R,I$) and near-IR
($J,H,K$) photometry of the X-ray source N2\_28 (upper) and the ERO
which lies a projected 12 arcsec away from it (lower). Also shown are
the best fitting reddened (solid) and unreddened (dashed) template
SEDs, which are all instantaneous burst models. For N2\_28 (upper) the
models shown have $z=0.99$, age\,=\,3.5\,Gyr, $A_V=0.6$ and $z=1.40$,
age\,=\,2.0\,Gyr, $A_V=0$. For the ERO near the X-ray source (lower)
the best fit models have $z=1.19$, age\,=\,2.3\,Gyr, $A_V=0.7$ and
$z=1.27$, age\,=\,5.5\,Gyr, $A_V=0$.}}

\end{figure}

\begin{figure} 
\vspace{0.5cm}
\epsfxsize=0.48\textwidth 
\epsfbox{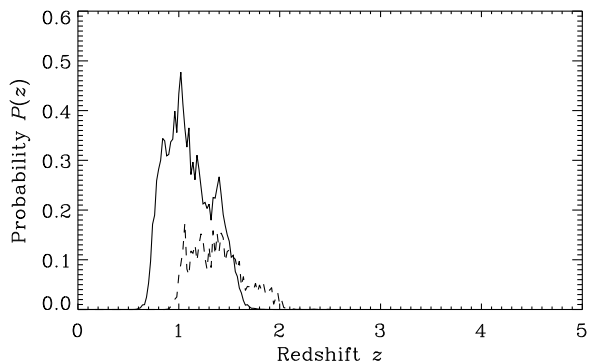} 
\vspace{0.6cm}
{\caption[junk]{\label{fig:chisq228} Probability distribution for the
redshifts of N2\_28 (solid) and the nearby ERO (dashed) from
{\footnotesize HYPERZ}.}}
\end{figure}

Fig.\ \ref{fig:sed228} shows the optical and near-IR photometry for
these two galaxies. Also shown are the best-fit reddened and
unreddened galaxy models. Fig.\ \ref{fig:chisq228} shows the
probability distributions of the redshift of these two galaxies for
the fits allowing reddening. Both distributions peak at a redshift of
about one. The 90\% confidence range for N2\_28 is $0.80 < z < 1.20$,
whilst that for the nearby ERO is $1.13 < z < 1.61$. The weighted mean
redshift for N2\_28 evaluated over all possible models (assuming equal
priors for all models) is $z=1.11$.  Given the evidence for clustering
of extremely red galaxies in this field it is quite probable that
these two galaxies are physically associated with redshifts of $z
\approx 1.2$. Although evolved (several Gyr) single burst models with
moderate amounts of reddening ($A_V \approx 0.5$) provide the best fit
to the data, unreddened models with similar ages also fit the data
well.

At redshifts between 1.06 and 1.25, the strong narrow emission lines
which one would expect in the near-IR such as H$\alpha$, H$\beta$ and
[O\,{\small III}] lie in the gaps either side of the $J$-band which
are not covered by the OHS spectra due to strong atmospheric
absorption. Therefore the lack of emission lines observed from the
X-ray source does not necessarily mean that it does not possess strong
emission lines. The most stringent limit on emission line luminosity
for an object at this redshift comes from the lack of [O\,{\small II}]
$\lambda 3727$ emission in the optical spectrum.  Given that the
expected wavelength of [O\,{\small II}] falls in regions affected by
strong sky emission lines, the limiting line flux is $<3
\times10^{-20}$ W m$^{-2}$, corresponding to a limiting luminosity of
$<2 \times 10^{34}$ W at $z=1.1$. Assuming an emission line ratio of
[O\,{\small III}]/[O\,{\small II}] = 3 for Seyfert 2s (Ferland \&
Netzer 1983) this limit is similar to that of [O\,{\small III}] on the
OHS spectrum if it is not in a gap ($1 \times 10^{-19}$ W~m$^{-2}$).
The limit on the emission line to hard X-ray luminosity for N2\_28
[$\log(L_{\rm [O\,{\small III}]}/L_{\rm X})<-1.5$] is greater than the
typical ratio for low redshift Seyfert 2s of [$\log(L_{\rm [O\,{\small
III}]}/L_{\rm X}]\approx-2$ (Mulchaey et al. 1994). Therefore this
X-ray source is not necessarily especially sub-luminous in emission
lines.

The X-ray luminosity of N2\_28 is much higher than can be accounted
for by star-formation (e.g. Brandt et al. 2001) and, in addition to
the X-ray hardness, indicates the presence of an obscured AGN at a
level just below that normally used to divide quasars from
Seyferts. The X-ray source resides in a host galaxy which has a
luminosity a factor of two lower than a passively evolving $L_{\star}$
galaxy assuming our redshift estimate is correct. Even if at $z=1.5$
the luminosity would only be equal to a passively evolving $L_{\star}$
galaxy. This host is much less luminous than $z \approx 1$ radio
galaxies which have a mean luminosity of $3L_{\star}$ (e.g. Willott et
al. 2002b) and probably reflects the fact that this source has a
smaller black hole and a lower accretion rate. If the nearby extremely
red galaxies are indeed members of a cluster containing the X-ray
source, then the galaxy hosting the X-ray source is not even the
dominant galaxy in the cluster.

\section{Conclusions}

The shape of the X-ray background requires the majority of sources in
hard X-ray selected surveys to have flat X-ray slopes, most likely due
to significant absorption of intrinsically steep slopes (e.g. Comastri
et al. 1995). The range of X-ray absorbing columns in AGN and its
evolution is poorly constrained at present and is one of the major
aims of the on-going \chandra and \xmm surveys. The finding of objects
such as N2\_25 which have only a small amount of optical reddening and
a much greater X-ray absorbing column further complicates the picture and
shows that optical obscuration does not always equal X-ray
absorption. Therefore the popular radio-loud unified scheme which
unifies radio galaxies and quasars on the basis of their orientation
(by means of optical obscuration -- Barthel 1989) may be quite
different from that of X-ray selected AGN which are classified by
their X-ray absorption. Unfortunately, investigating the links between
the radio-loud and radio-quiet unified schemes is further complicated
by the fact that luminous radio sources emit X-rays associated with
the radio emission. As may be the case for the radio galaxy N2\_21,
this emission could dominate over that of the absorbed nucleus,
meaning that even with the high spatial resolution of \chandra, it is
difficult to accurately measure the nuclear absorbing columns in
powerful radio galaxies.

Little is known about the large numbers of optically-faint ($R>25$)
sources contained within hard X-ray surveys, due to the fact that they
are too faint for optical spectroscopy. As discussed by Alexander et
al. (2001) their X-ray to optical flux ratios, inferred luminosities
(given their likely redshifts at $z>1$) and variability suggest that
most are AGN. Since some of these objects are extremely red and hence
accessible with near-IR spectroscopy on 8-m class telescopes, we have
attempted to determine the redshifts and nature of two such sources
with Subaru Telescope. Emission lines were only detected from the
radio-loud source N2\_21, although the lack of lines in the spectrum
of N2\_28 could plausibly be due to the strongest lines falling in the
gap between the $J$ and $H$ bands. The upper limit on the ratio of
emission line to hard X-ray luminosity for N2\_28 is greater than the
typical ratio for low redshift Seyfert 2s, so there is no evidence
that such sources have particularly weak emission lines.

It seems likely that most optically-faint hard X-ray sources are
obscured high-redshift AGN in which the rest-frame UV/optical light is
dominated by the host galaxies. This provides an opportunity to study
the nature of high-redshift galaxies with the only selection effect
being that they contain a massive black hole which is undergoing
substantial accretion. Our observations of the radio galaxy N2\_21
show that its luminous host galaxy contains a large amount of dust and
is likely to be undergoing a massive starburst. In contrast, the
sub-$L_{\star}$ host of the radio-quiet N2\_28 has an SED consistent
with an evolved stellar population at $z\approx 1$ and may be located
in a cluster. With the on-going follow-up observations of the EDXS, we
plan to characterize the properties of a larger sample of
optically-faint hard X-ray hosts and shed further light on the nature
of the sources responsible for the hard X-ray background.

\section*{Acknowledgements}

Thanks to Dave Alexander and Ross McLure for useful discussions. Many
thanks to the referee Dan Stern for his interesting suggestions.
Thanks to Katherine Blundell for help with obtaining the 8.4-GHz radio
data. Based on data collected at Subaru Telescope, which is operated
by the National Astronomical Observatory of Japan. The William
Herschel Telescope is operated on the island of La Palma by the Isaac
Newton Group in the Spanish Observatorio del Roque de los Muchachos of
the Instituto de Astrofisica de Canarias. The United Kingdom Infrared
Telescope is operated by the Joint Astronomy Centre on behalf of the
U.K. Particle Physics and Astronomy Research Council. The VLA is a
facility of the National Radio Astronomy Observatory (NRAO), which is
operated by Associated Universities, Inc. under a cooperative
agreement with the National Science Foundation. CJW thanks PPARC for
support.

\end{document}